\documentclass{article}
\usepackage{dcase2021,amsmath,graphicx,url,times,booktabs,tabularx,amsfonts,siunitx,microtype}
\usepackage[numbers]{natbib}
\usepackage{microtype}

\let\OLDthebibliography\thebibliography
\renewcommand\thebibliography[1]{
  \OLDthebibliography{#1}
  \setlength{\parskip}{0pt}
  \setlength{\itemsep}{4.4pt plus 0.1ex}
}

\title{CL4AC: A Contrastive Loss for Audio Captioning}

\name{
      Xubo Liu$^{1*}$,
      Qiushi Huang$^{2,3}\sthanks{The first two authors contributed equally to this work.}$,
      Xinhao Mei$^{1}$,
      Tom Ko$^{3}$, 
      H Lilian Tang$^{2}$, 
      Mark D. Plumbley$^{1}$,
      Wenwu Wang$^{1}$
      }
\address{$^1$ Centre for Vision, Speech and Signal Processing (CVSSP), University of Surrey, UK,\\
\{xubo.liu, x.mei, m.plumbley, w.wang\}@surrey.ac.uk\\          
        $^2$ Department of Computer Science, University of Surrey, UK, \{qiushi.huang, h.tang\}@surrey.ac.uk\\ 
        $^3$ Southern University of Science and Technology, Shenzhen, China, tomkocse@gmail.com\\ 
 }

\begin{document}

\ninept
\maketitle

\begin{sloppy}

\begin{abstract}
Automated Audio captioning (AAC) is a cross-modal translation task that aims to use natural language to describe the content of an audio clip. As shown in the submissions received for Task 6 of the DCASE 2021 Challenges, this problem has received increasing interest in the community. The existing AAC systems are usually based on an encoder-decoder architecture, where the audio signal is encoded into a latent representation, and aligned with its corresponding text descriptions, then a decoder is used to generate the captions. However, training of an AAC system often encounters the problem of data scarcity, which may lead to inaccurate representation and audio-text alignment. To address this problem, we propose a novel encoder-decoder framework called \textbf{C}ontrastive \textbf{L}oss for \textbf{A}udio \textbf{C}aptioning (CL4AC). In CL4AC, the self-supervision signals derived from the original audio-text paired data are used to exploit the correspondences between audio and texts by contrasting samples, which can improve the quality of latent representation and the alignment between audio and texts, while trained with limited data. Experiments are performed on the Clotho dataset to show the effectiveness of our proposed approach.


\end{abstract}


\section{Introduction}
\label{sec:intro}
Automated Audio captioning (AAC) is a cross-modal translation task of generating a natural language description for an audio clip. It has various potential applications. For example, AAC can be used for generating subtitles for the audio content in a television program, or for generating text descriptions of audio to help the hearing impaired in accessing audio content. It can also be used by sound search engines to achieve more accurate retrieval and recommendation, or by a surveillance system to facilitate the detection of acoustic anomalies. The AAC problem has attracted increasing interest from the acoustic signal processing and machine learning communities in recent years.

 Existing AAC systems are usually based on an encoder-decoder architecture \cite{drossos2017automated, chen2020audio, koizumi2020transformer, tran2020wavetransformer, mei2021audio}. The audio data is encoded into a latent representation and aligned with its corresponding text description. Then a decoder is used to generate the captions. Training of an AAC system often encounters the problem of data scarcity, which may lead to inaccurate representation and audio-text alignment. For example, Clotho \cite{drossos2020clotho} is a popular AAC dataset and was used for the DCASE challenge. However, it only contains 6974 audio samples, and each audio sample has five captions. To address this problem, information from keywords has been exploited for AAC \cite{koizumi2020transformer, takeuchi2020effects, eren2020audio}. The keywords of the caption are tagged firstly and then used to assist the generation of captions. However, due to the diversity of keywords, the tagging results of unseen audio clips may not be accurate in the inference stage. On the other hand, transfer learning techniques \cite{xinhao2021_t6, yuan2021_t6} have been widely used in task 6 of the DCASE 2021 challenge, offering substantially improved performance. However, transfer learning relies heavily on large-scale external data \cite{kim2019audiocaps} and pre-trained models \cite{kong2020panns}.

Contrastive learning \cite{CPC, supervised_contrastive_learning} is a self-supervised paradigm that helps the model obtain high-quality representation. Inspired by the recent success of contrastive learning in computer vision (CV) \cite{chen2020simple} and natural language processing (NLP) \cite{Gunel2020SupervisedCL, huang2021tokenlevel}, we propose a novel encoder-decoder framework called \textbf{C}ontrastive \textbf{L}oss for \textbf{A}udio \textbf{C}aptioning (CL4AC). In CL4AC, the self-supervision signals derived from the original audio-text paired data are used to exploit the correspondences between audio and texts by contrasting samples. More precisely, we construct mismatched audio-text pairs as negative samples. Then, a contrastive learning objective is designed to maximize the difference between the representation of the matched audio-caption pair derived from the negative pairs. In this way, the quality of latent representation and the alignment between audio and texts can be improved without introducing large-scale external data, when they are trained with limited amount of data. To the best of our knowledge, contrastive learning approach has not been used for AAC in the literature. 



The remainder of this paper are organised as follows. We introduce our proposed CL4AC in Section \ref{sec:framework}. Experiments are described in Section \ref{sec:experiments}. Results are shown in Section \ref{sec:results}. Finally, we conclude our work and discuss the future work in Section \ref{sec:conclusion}. The code of this work is made available on GitHub\footnote{\url{https://github.com/liuxubo717/cl4ac}}.


\section{Contrastive Loss for Audio Captioning}
\label{sec:framework}

In this section, we present our proposed contrastive learning framework for audio captioning (CL4AC). We first introduce the encoder-decoder architecture of CL4AC in Section 2.1. Then, we present the contrastive learning framework in Section 2.2.

\subsection{Encoder-Decoder architecture}
We first define the notations used in this section. The training data for AAC consists of paired audio and texts data. We denote a training set of $N$ audio-text pairs by $D=\{(a_n, C_n)\}_{n=1}^N$, where $a \in \mathbb R^{H\times W}$ is the log mel-spectrogram of an audio clip with $H$ and $W$ being its height and width, respectively, $C=\{w_m\}_{m=1}^M$ is the token sequence of a caption where $w_m$ is the $m$-th token in the caption $C$ having $M$ tokens, $a_n$ is the log mel-spectrogram of the $n$-th audio clip in the dataset, and $C_n$ is the token sequence of the $n$-th caption in the dataset. 

The sequence-to-sequence architecture with Convolutional Neural Network (CNN) encoder and Transformer decoder are used as the basis of our proposed framework, as shown in Figure 1. This architecture was shown to offer the state-of-the-art performance \cite{xinhao2021_t6, yuan2021_t6} in Task 6 of the DCASE 2021 challenge. 

\begin{figure}[!ht]
    \centering
    \includegraphics[width=\columnwidth]{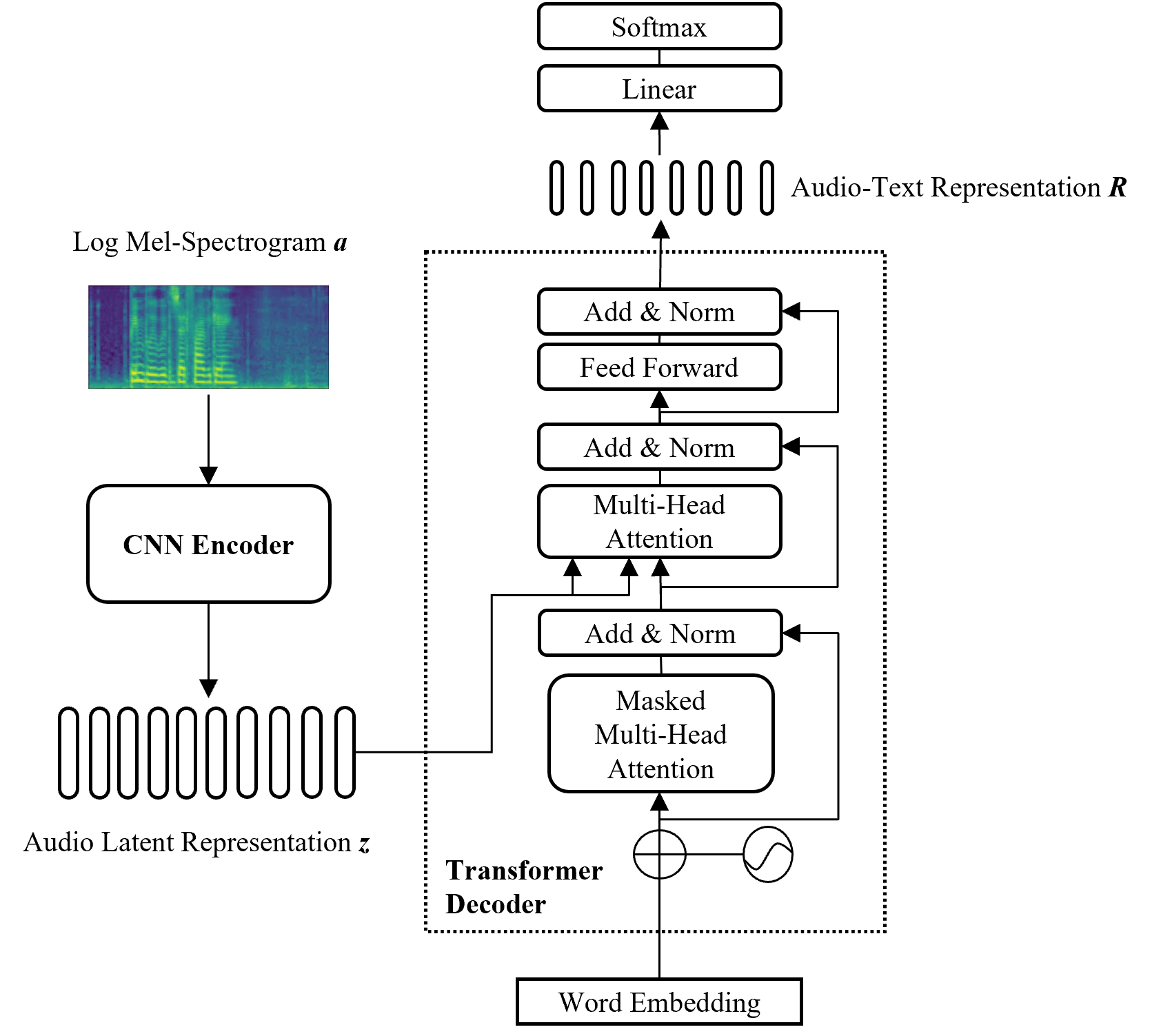}
    \caption{Sequence-to-sequence architecture with CNN encoder and Transformer decoder for audio captioning. The components in the dashed box indicate the Transformer decoder.}
    \label{fig:network}
\end{figure}
\subsubsection{CNN encoder}
Pre-trained audio neural networks (PANNs) \cite{kong2020panns} have demonstrated a powerful ability in extracting latent representation of audio signals for different downstream audio recognition tasks. To benefit from its high-quality audio representation, we choose PANNs as the encoder, which will be described in Section 3.3 in details. The PANNs encoder takes the log mel-spectrogram $a$ of an audio clip as the input and extracts its latent representation $z \in \mathbb R^{H' \times W'}$. Formally:
\begin{equation}
  \label{eqn:encoder}
  z = \operatorname{Encoder}(a).
\end{equation}

\subsubsection{Transformer decoder}

The Transformer model has shown the state-of-the-art performance on language-related cross-modal task \cite{oscar, cornia2020meshed}, and is used as the decoder in our work. There are two main components in the decoder. Firstly, each token $w_m$ in the input token sequence $C$ is converted into a word embedding $e_m \in \mathbb{R}^{1\times E}$, where $E$ is the dimension of the word embedding, by the word2vec algorithm using Continuous Bag of Words Model (CBOW) \cite{CBOW} and Skip-Gram \cite{SKIPGRAM} model trained purely on the caption corpus. Then the word embedding of tokens are fed into the first self-attention layer to obtain their hidden states. The latent representation $z$ of an audio clip extracted by the encoder is aligned and calculated with the hidden states of tokens, then the audio-text representation is obtained by the transformer decoder, denoted as $R \in \mathbb{R}^{M\times T}$, which consists of $M$ vectors $\{r_m\}_{m=1}^M$, where the number of vectors is equal to the length of the input token sequence $C$ and the dimension of each vector is $T$. The vector $r_m$ of the audio-text representation $R$ is calculated based on the word embeddings $\{e_1, ..., e_{m-1}\}$ and the audio latent representation. Hence, each $r_m$ corresponds to the token $w_m$ in the input token sequence $C$ one-to-one, which can be used to predict the probability of the word over the vocabulary after it is passed through the final linear layer with softmax function. The transformer decoder predicts the $m$-th word $w_m$ based on the previous tokens $\{w_1,...,w_{m-1}\}$ and the audio latent representation $z$, as follows, 


\begin{equation}
  \label{eqn:decoder}
  p(w_m|z,w_1,...,w_{m-1}) = \operatorname{Decoder}(z,w_1,...,w_{m-1}).
\end{equation}
The training objective is to optimize the cross entropy (CE) loss defined in terms of the predicted words as:
\begin{equation}
  \label{eqn:decoder}
  \operatorname{Loss_{CE}} = -\mathbb E_{(a,C)\sim D}\operatorname{log}p(w_m|z,w_1,...,w_{m-1}).
\end{equation}

\subsection{Contrastive learning framework}
To obtain accurate audio-text representation $R$ while the model is trained with limited data, we use the self-supervised signal derived from the audio-text training data by contrasting samples. First, we construct mismatched audio-text pairs as negative samples. Then, a contrasting auxiliary task is designed to maximize the difference between the representation $R$ of the matched audio-text pair derived from negative pairs. The representations of the audio-text paired data are pulled together in the latent space while simultaneously pushing apart clusters of unpaired negative data by contrastive learning, as shown in Figure 2. In this way, the quality of audio-text representation and the alignment between audio and texts can be improved.
\begin{figure}[!ht]
    \centering
    \includegraphics[width=\columnwidth]{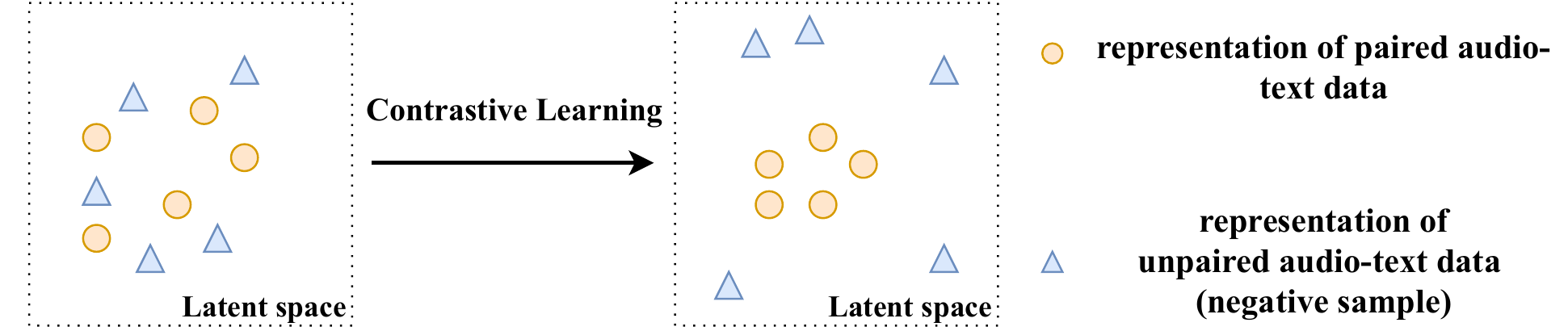}
    \caption{The representations of the audio-text paired data are pulled together in the latent space while simultaneously pushing apart clusters of unpaired negative data by Contrastive Learning (CL).}
    \label{fig:network}
\end{figure}

\begin{figure*}
  \centering
  \includegraphics[width=\linewidth]{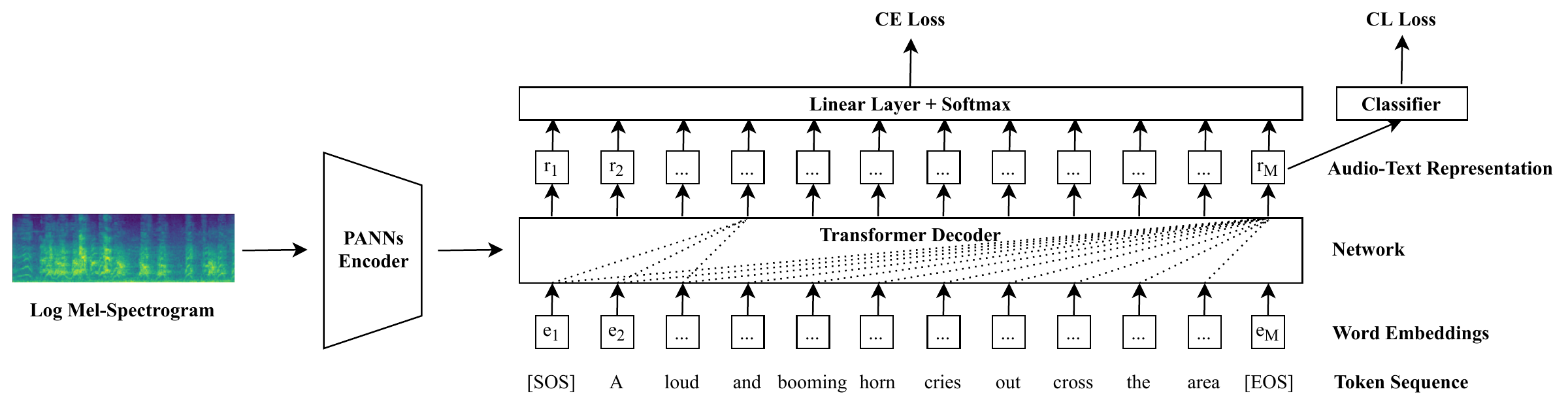}
  \caption{Contrastive loss for audio captioning (CL4AC) framework. The dashed lines indicate that the vector $r_m$ of the audio-text representation $R$ is calculated based on the word embeddings $\{e_1, ..., e_{m-1}\}$ and the audio latent representation obtained from PANNs. The last audio-text representation vector $r_M$ is fed to the classifier $f(.)$ whose output is used to calculate the Contrastive Learning (CL) loss. }
  \label{fig:vqvae}  
\end{figure*}

\begin{table*}[ht]
\centering
\begin{tabular}[\linewidth]{c | c | c} 
 \hline
  Example & paired caption $C$ & unpaired caption $C_{negative}$ \\ 
 \hline
  & Something goes round that is playing its song & The Air is blowing some what fast outside \\
 & At the fair, music is playing near a carousel through the speaker & A hand held sander was used as various speeds  \\
 audio $a$ & Chiming of bells, whistles and horns at a performance & A hard gravel ground is walked on by someone \\
 & Fair kind music is being played at the circus grounds & A person using a hard object to tap and scrape glasses \\
 & Polka or fair kind of music is being played & The wind is blowing and the waves are flowing \\
 \hline
\end{tabular}

\caption{Examples of paired audio-text training data  $x=(a,C)$ and negative training sample  $x_{negative}=(a,C_{negative})$. Examples are selected from the Clotho dataset, where each audio data has five corresponding captions.}.
\label{table:tab_caps} 
\end{table*}
More specifically, for each anchor audio-text paired training data $x=(a, C)$, we replace the caption $C$ by $C_{negative}$ which is a randomly selected caption unpaired with $a$ in the training set $D$. Then, the mismatched audio-text pair as the negative training sample is constructed, denoted as $x_{negative}=(a,C_{negative})$. Table 1 shows the examples of $x$ and $x_{negative}$ in the Clotho dataset. Since the last vector in the audio-text representation $R$ is able to attend the context of all input tokens and the audio feature, the value of last vector of $R$ is fed into a binary classifier $f(.)$ to predict whether the input audio and text data are paired $(y=0)$ or not $(y=1)$. The contrastive learning (CL) loss for this auxiliary task is defined as follows:


\begin{equation}
  \label{eqn:decoder}
  \operatorname{Loss_{CL}} = -\mathbb E_{x'\sim D'}\operatorname{log}p(y|f(x')),
\end{equation}
where $D'$ is the extended training set by merging the negative samples into the original training set $D$ and $x'$ is the audio-text pair drawn from $D'$. The full training objective of CL4AC is:
\begin{equation}
  \label{eqn:decoder}
  \operatorname{Loss_{Training}} = (1-y)\operatorname{Loss_{CE}} + \operatorname{Loss_{CL}}. 
\end{equation}
When the input is a negative audio-text pair, the gradient provided by the CE loss is meaningless, for this case, only CL loss is used for updating the model. The framework of CL4AC is shown in Figure 3.

\section{Experiments}
\label{sec:experiments}
\subsection{Dataset}
Clotho \cite{drossos2020clotho} is an AAC dataset whose sound clips are from the Freesound platform and annotated by Amazon Mechanical Turk. Clotho v2 was released for Task 6 of the DCASE 2021 Challenge, which contains \num{3839}, \num{1045} and \num{1045} audio clips for the development, validation and evaluation split respectively. The sampling rate of all audio clips in Clotho dataset is \SI{44100}{\Hz}. Each audio clip has five captions. Audio clips are of \num{15} to \num{30}s duration and captions are eight to \num{20} words long. We merge the development and validation split, forming a new training set with \num{4884} audio clips. The performance of AAC system is evaluated on the evaluation split.

\subsection{Data pre-processing}
We use the original sampling rate to load audio data, and an \num{64}-dimensional log mel-spectrogram is calculated using the short-time Fourier transform (STFT) with a frame size of \num{1024} samples, a hop size of \num{512} samples, and a Hanning window. SpecAugment \cite{park2019specaugment} is used for data augmentation.
 
We transform all captions in the Clotho dataset to lower case with punctuation removed. Two special tokens ``\texttt{\textless sos\textgreater}'' and ``\texttt{\textless eos\textgreater}'' are added on the start and end of each caption. The vocabulary of the Clotho dataset contains \num{4367} words.

\subsection{Model implementation}
CNN-10 of PANNs \cite{kong2020panns} is used as the encoder to prevent over-fitting while trained with limited data. Specifically, the CNN-10 consists of four convolutional blocks where each has two convolutional layers with a kernel size of $3 \times 3$. Batch normalization and ReLU are used after each convolutional layer. The channels number of each block are \num{64}, \num{128}, \num{256} and \num{512}, respectively. An average pooling layer with kernel size $2 \times 2$ is applied between them for down-sampling. Global average pooling is applied along the frequency axis after the last convolutional block followed by two fully connected layers to align the dimension of the output with the decoder input. Two transformer blocks with four heads and \num{128} hidden units are used as the decoder. The implementation for the encoder and decoder is the same as that in our DCASE 2021 Challenge system\footnote{\url{https://github.com/XinhaoMei/DCASE2021_task6_v2}}, which is the highest-scoring system without using model ensembles.

We trained the proposed model using Adam \cite{kingma2014adam} optimizer with a batch size of \num{16}. Warm-up is used in the first \num{5} epochs to increase the learning rate to the initial learning rate linearly. The learning rate is then decreased to \num{1/10} of itself every \num{10} epochs. Dropout with a rate of \num{0.2} is applied in the proposed model to mitigate the over-fitting problem. We train the model for 30 epochs with an initial learning rate of \num{5e-4} on the training set of the Clotho dataset. 
\begin{table*}[!t]
\centering
\begin{tabular}[\linewidth]{c c c c c c c c c c} 
 \hline
 Model & BLEU$_{1}$ & BLEU$_{2}$ & BLEU$_{3}$ & BLEU$_{4}$ & ROUGE$_{L}$ & METERO & CIDEr & SPICE & SPIDEr \\ 
 \hline
 Baseline & 0.550 & 0.345 & 0.222 & 0.139 & 0.372 & 0.169 & 0.356 & 0.115 & 0.235 \\
 CL4AC & 0.553 & 0.349 & 0.226 & 0.143 & 0.374 & 0.168 & 0.368 & 0.115 & 0.242 \\
 \hline
\end{tabular}

\caption{Performance of models is evaluated on the Clotho v2 evaluation set. Baseline: baseline system described in Section 3.4, which is similar to our DCASE submitted system but without transfer learning and reinforcement learning techniques. CL4AC: Proposed framework Contrastive Loss for Audio Captioning (CL4AC). During the inference stage, captions are generated using greedy search.}
\label{table:tab_results} 
\end{table*}
\subsection{Baseline system}
The baseline system is similar to our DCASE 2021 system which uses transfer learning (TL) from external dataset and reinforcement learning (RL) \cite{xinhao2021_t6}. Our motivation is to mitigate the data scarcity problem for AAC without introducing external datasets, so we train the baseline without using the TL technique. Previous studies \cite{mei2021encoder, dialogpt} proved that although RL techniques can optimize neural networks towards non-differentiable metrics, they may generate syntactically incorrect and incomplete captions. Thus, RL is also removed in the baseline system. The hyper-parameters used for training the baseline system are similar to the proposed model (as described in Section 3.3), except that the training batch size is 32 and the initial learning rate is \num{1e-3}.

\subsection{Evaluation}

During the inference stage, the mel-spectrogram of an audio clip along with the special token ``\texttt{\textless sos\textgreater}'' are fed into the encoder and decoder separately to generate the first token. Afterwards, the following tokens are predicted in terms of the previously generated tokens until the token ``\texttt{\textless eos\textgreater}'' or the maximum length (35 words in our experiments) is reached. The greedy search strategy is used to generate captions.

We evaluate the performance of the proposed framework using the same metrics adopted in Task 6 of the DCASE 2021 Challenge, including machine translation metrics: BLEU$_{n}$ \cite{papineni2002bleu}, METEOR \cite{lavie2007meteor}, ROUGE$_{l}$ \cite{lin2004rouge} and captioning metrics: CIDEr \cite{vedantam2015cider}, SPICE \cite{anderson2016spice}, SPIDEr \cite{liu2017improved}. BLEU$_n$ measures the quality of the generated text by calculating the precision of $n$-gram inside the text, which is an inexpensive metric to measure the correspondence between generated text and the ground truth. Generally, the higher BLEU$_n$ usually implies better precision and fluent text. The SPIDEr, a combination of SPICE and CIDEr, is designed for image captioning task measurement, which considers scene graph inside the generated caption and the term frequency-inverse document frequency (TF-IDF) of the $n$-gram. By considering the scene graph and the TF-IDF of $n$-gram, the metric will focus on the relationships among objects and the text's property, which ensures the semantic fidelity to the audio and the syntactical fluency of the language.

\section{Results}
Table \ref{table:tab_results} shows the performance of our proposed method on the Clotho v2 evaluation set. By adopting the contrastive loss technique during the training process, all the metrics except METERO increased on the evaluation set. For BLEU$_1$, BLEU$_2$, BLEU$_3$, BLEU$_4$, the relative improvement percentages for contrastive loss are 0.55\%, 1.16\%, 1.80\%, and 2.88\%, respectively. The $n$ in BLEU$_n$ means the $n$-grams matching between the predicted results and ground truths. The ascending increases of the relative improvement from BLEU$_1$ to BLEU$_4$ show that our proposed method generates more matching $n$-grams, demonstrating a more fluent and better quality captioning result. Besides, CIDEr and SPIDEr, the captioning metrics, obtained 3.37\% and 2.98\% relative improvement correspondingly. The better CIDEr and SPIDEr ensure the captions are better semantically faithful to the audio clip with the better language fluency. Numerical improvement of the machine translation and captioning metrics shows the effectiveness of CL4AC while trained with limited data.

\label{sec:results}

\section{Conclusions}
\label{sec:conclusion}
This paper demonstrated the problem of data scarcity for AAC, which may lead to the inaccurate representation and audio-text alignment. To alleviate this issue, a novel encoder-decoder framework called \textbf{C}ontrastive \textbf{L}oss for \textbf{A}udio \textbf{C}aptioning (CL4AC) was proposed to learn a better cross-modal representation. In CL4AC, the self-supervision signals derived from the original audio-text data are used to exploit the correspondences between audio and text by contrasting samples in a limited dataset setting. Experiment results on BELU${_n}$, CIDEr, and SPIDEr showed the effectiveness of the proposed approach with a relative improvement of up to 3.37\%, compared to the baseline system. 
In future work, we will explore more contrastive learning approaches for AAC, such as Momentum Contrast (MoCo) \cite{MoCo} and SimCLR \cite{chen2020simple}.

\section{ACKNOWLEDGMENT}
\label{sec:ack}
This work is partly supported by grant EP/T019751/1 from the Engineering and Physical Sciences Research Council (EPSRC), a Newton Institutional Links Award from the British Council, titled ``Automated Captioning of Image and Audio for Visually and Hearing Impaired" (Grant number 623805725) and a Research Scholarship from the China Scholarship Council (CSC) No. 202006470010. 

\bibliographystyle{IEEEtran}
\bibliography{refs}

%
%
%
%
%
%
%
%
%

\end{sloppy}
\end{document}